# A time of flight method to measure the speed of sound using a stereo sound card

*Carlos C. Carvalho*, E.S. Garcia de Orta, Porto, 4150-620 Porto, Portugal

*J. M. B. Lopes dos Santos*, CFP e Departamento de Física, Faculdade de Ciências Universidade do Porto, 4169-007 Porto, Portugal

*M. B. Marques*, Departamento de Física, Faculdade de Ciências Universidade do Porto, 4169-007 Porto, Portugal

Most homes in developed countries possess a sophisticated data acquisition board, namely the PC sound board. Designed to be able to reproduce CD quality stereo sound, it must have a sampling rate of at least 44 kHz, and very accurate timing between the two stereo channels. With a very simple adaptation of a pair regular PC-microphones, a computer with a stereo sound board, and sound analysis software, we were able to implement a method of accurate measurement of the speed of sound with several appealing features:

a)  The most expensive equipments, the PC, the sound board and sound analysis software are available in many homes and almost all schools; the two microphones and the additional electronics (see below) can be acquired for less than 50 $.
b)  The concept of the experiment is sufficiently simple to be grasped by very young students, 13~14 years old.
c)  The experiment itself is so straightforward, that in a recent *Open Day* at our department every group of students passing through the experiment was able to complete, in a couple of minutes, several measurements of $c_s$, the speed of sound in air.
d)  It is possible to use the same apparatus to measure the speed of sound in solids.

Measurement of the speed of sound is a very common procedure in many schools, and many appealing methods have been described in the literature. Most use standing waves, exploring resonance phenomena, or the spatial variation of the phase[1-3]. Less common are methods based on the time of flight of a pulse, and these most often involve triggering of the measuring oscilloscope with the signal that generates the sound pulse and timing the time delay of the pulse picked up by a conveniently placed microphone[4-5]. Loren Winters has reported a method similar in principle to the present one, but which uses a completely different detection system[6].

## Principle

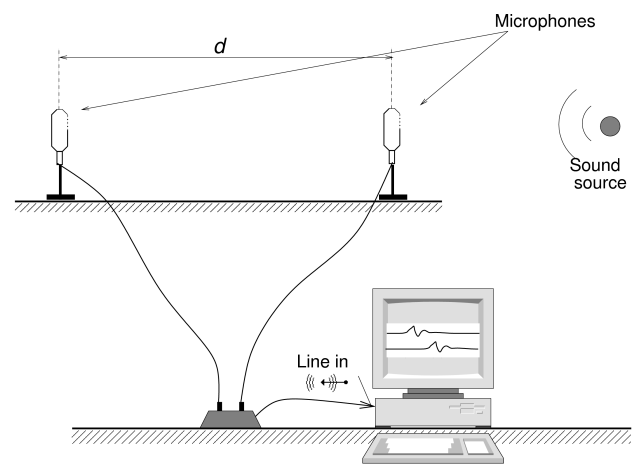

*Figure 1: By feeding the signals from each microphone separately to the L and R stereo channels and measuring the time delay between them using sound analysis software, we can measure the time of travel of a sound pulse between the two microphones.*

The basic principle of this experiment is illustrated in Figure 1. We use two PC-microphones separated by a known distance, $d$, and feed the signal from each microphone to a separate sound channel of the sound board using the *line-in* connection. A sharp sound (clapping hands, or slamming two flat pieces of wood together are very adequate sound sources) is produced at a point along the line that joins the two microphones. The sound analysis software can display the signals recorded in the left ($L$) and right ($R$) channels. The signal recorded by the microphone farthest from the source shows a time delay $\Delta t = d/c_s$ relative to the other one, from which we can determine $c_s$.

## Apparatus

The most common PC-microphones require power, usually drawn the *micro* connection of the PC sound board. To feed the signal of each microphone to separate stereo channels, we needed to use the *line-in* entrance to the sound board, which does not provide power. There are two alternatives to overcome this difficulty. Self powered microphones can be used with an adapter from two mono sockets to a stereo plug. Note that this is not the much used power splitter that allows the connection of two headphones to a single output, but rather a special adapter that connects each input mono channel socket to only one of the two stereo channels in the output plug; it can be acquired in electronics hardware shops for about 4 $. The alternative is build a very simple electronics box, with two 1.5 V batteries, for which we give details in the Appendix. In this case one can use conventional, run-of-the-mill, PC microphones.

For sound analysis software we used *Cool-Edit Pro©*. This program is no longer available as shareware. We also tested *Audacity©* (http://audacity.sourceforge.net/, a cross-platform freeware program, with very good results. Two screen shots of the data acquired with *Cool-Edit Pro©* during one measurement are shown in Fig. 2. In the top panel, with lower time resolution, two sharp pulses are seen, apparently simultaneous. Zooming in on the time axis, though, one clearly sees the time delay between the two signals (lower panel). Using the cursor, the region between the starts of each of the pulses has been selected, and the corresponding duration (in samples) is shown by the software. Knowing the sample rate (in our case $f = 44100\,Hz$), one can easily determine the time of travel of the sound pulse between the two microphones.

The software takes about 130 samples in the time that sound travels $1\,m$ for a sampling interval of $\delta t = 1/f = 0.023\,ms$, ; the time of travel of the sound pulse between microphones can be measured with 1% precision without difficulty.

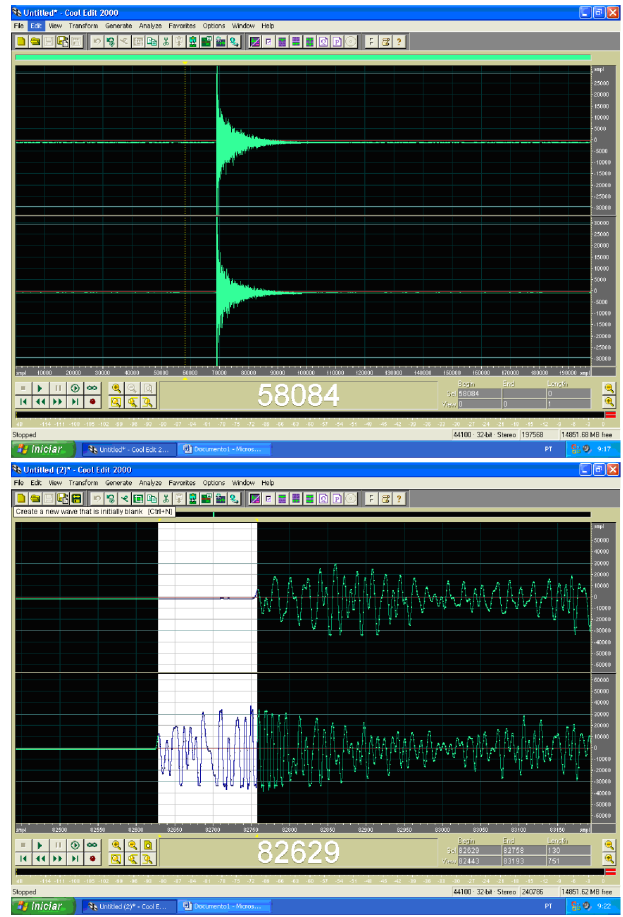

*Figure 2: Two screen shots of a measurement of a single sound pulse: in the top panel, with low time resolution, the signals in both stereo channels appear simultaneous. Zooming in on the time scale, one can effectively see a time delay. The time interval highlighted in white is given in samples in lower right*

## Results

Although it is possible to obtain a very reasonable value of $c_s$ in a single measurement, a more accurate determination is achieved by varying the distance between the two microphones, and by plotting the distance $d$ as a function of the time delay between the two channels (see Fig. 3). The slope is directly the sound speed. The conceptual value of this procedure is even more important than the added precision. By noting that the measured delay is proportional to the distance between the microphones, the students conclude that the observed time delay is, in fact, the time of travel of the sound pulse between the microphones.

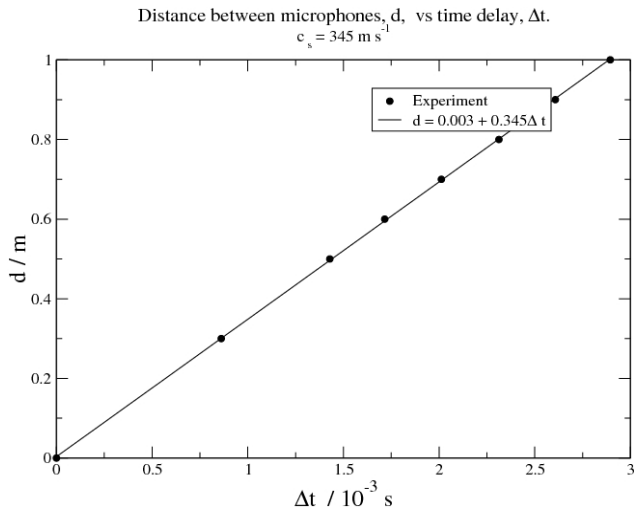

*Figure 3: Results of a measurement of the speed of sound. The error bars in and in are no larger than the size of the dots.*

By scotch-taping the microphones to a table top or a metal bar, we were able to detect sound pulses propagating in a solid, after tapping near one extremity. One only detects vibrations in which the surface of the solid moves in the direction of its normal. In solids, we can have transverse, as well as longitudinal, sound waves, with very different speeds. In either case, the speed of sound is much higher than in air, and, unless the microphones are very close, reflections and echoes become important; the identification of the time delay becomes somewhat problematic. For such short distances, the time delays start to become comparable to the time interval between samples, $\delta t = 0.023 \, ms$, and precise measurements are not possible. Still, we have been able to obtain very reasonable estimates of the speed of sound waves in solids with this method.

## Conclusions

We have presented an inexpensive apparatus for measuring the speed of sound, using a computer with a stereo sound board. It can serve as a very effective demonstration, providing a quick measurement of the speed of sound in air; we have used it with great success in *Open Days* in our Department. It can also be used for a full fledged laboratory determination of the speed of sound in air.

## Appendix

The electronics box that we used to connect conventional PC microphones to separate stereo channels of the *line-in* input can be easily assembled by any electronics hobbyist. The circuit scheme is shown in Fig. 4.

A small plastic box (obtainable in any electric supplies store) contains two 1.5 V batteries and two standard 3.5 mm female jacks, to which one connects the male connectors of the microphones. A three-wire cable, with a male 3.5 mm jack, connects the box to the PC *line-in* entry. The two $1 \mu F$ capacitors remove the DC component from the signal; the 2.2 kΩ load resistors are required to fix the DC set point and gain of the external circuit. We have used this assembly for more than two years without having to replace the batteries.

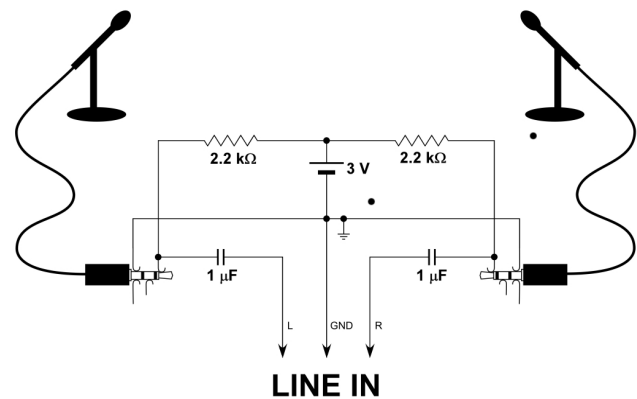

*Figure 4: Diagram of the circuit used to feed the signal from each microphone to a single stereo channel of the sound board. The resistors, capacitors, and batteries are enclosed in a small plastic box with two female 3.5 mm jacks, shown in the figure with the microphone connectors plugged in. A three-wire cable with a 3.5 mm male connector—identical to the two shown in the figure —plugs into the line-in entry of the sound board. The ground line (G) is attached to the connector at the base of the male jack (not shown); the left (L) and right (R)*

## Acknowledgments

We acknowledge the financial support of Fundação Calouste Gulbenkian, through Projecto Faraday, and thank Mr. Ernesto Pinheiro for his technical assistance with the electronics.

*Carlos Carvalho* is a Physics graduate of the University of Porto. Started his career as teaching assistant at the same University and, at present, teaches in secondary school.

*João M. B. Lopes dos Santos* is Associate Professor of Physics at the University of Porto: jlsantos@fc.up.pt.

*Manuel B Marques is* Assistant Professor of Physics at the University of Porto: mbmarque@fc.up.pt.

The authors recently collaborated in Project Faraday, a project involving five secondary schools, aimed at improving physics teaching at secondary level.